# End-User Development for Artificial Intelligence: A Systematic Literature Review

Andrea Esposito[✉], Miriana Calvano, Antonio Curci, Giuseppe Desolda, Rosa Lanzilotti, Claudia Lorusso and Antonio Piccinno

Department of Computer Science, University of Bari Aldo Moro, Bari, Italy
{andrea.esposito,miriana.calvano,antonio.curci,giuseppe.desolda,
rosa.lanzilotti,antonio.piccinno}@uniba.it,
c.lorusso36@studenti.uniba.it

**Abstract.** In recent years, Artificial Intelligence has become more and more relevant in our society. Creating AI systems is almost always the prerogative of IT and AI experts. However, users may need to create intelligent solutions tailored to their specific needs. In this way, AI systems can be enhanced if new approaches are devised to allow non-technical users to be directly involved in the definition and personalization of AI technologies. End-User Development (EUD) can provide a solution to these problems, allowing people to create, customize, or adapt AI-based systems to their own needs. This paper presents a systematic literature review that aims to shed the light on the current landscape of EUD for AI systems, i.e., how users, even without skills in AI and/or programming, can customize the AI behavior to their needs. This study also discusses the current challenges of EUD for AI, the potential benefits, and the future implications of integrating EUD into the overall AI development process.

## 1 Introduction

A very recent survey by the McKinsey Global Institute found that by 2022, approximately 50% of the surveyed companies will use AI in at least one function [1]. Furthermore, the recent proliferation of AI products such as Chat-GPT has contributed to the growing popularity of the topic, as evidenced by a possible correlation between the popularity of the two keywords in Google searches [2].

We would like to emphasize that in this paper, unless otherwise specified, we assume a broad definition of AI that includes autonomous systems using machine learning, neural networks, and statistical methods, as well as recommender systems, adaptive systems, and systems for face, image, speech, and pattern recognition.

The motivation for this study lies in the fact that the consequences of the "one size fits all" approach often adopted by AI systems can be an advantage when it comes to reaching a broader audience, thanks to its intrinsic generality. However, such an approach often renders the overall system inadequate for the specific and situational needs of different users. The full benefits of AI systems can be increased if new approaches are devised to allow non-technical users to be directly involved in the definition and personalization of AI technologies. In this direction, End-User Development (EUD) can provide a solution to these problems, allowing users to customize AI-based systems to their own needs and providing ways to deal with outliers and reduce bias. Another important motivation is that a few literature reviews deal with the topic of EUD for AI. For example, Gresse von Wangenheim et al. provide a comprehensive mapping of tools aiding the teaching of machine learning using visual programming paradigms [3]. Similarly, Hauck et al. provide an overview of the available tools that, using node- or block-based programming, allow the development of smart IoT devices (thus, powered by AI) [4]. An additional list of tools is provided by Queiroz et al., who provide a review of tools that may help teach lay people with a minimum understanding of what AI is [5]. More in line with our study, Li et al. provide a review of no/low code tools for AI ; however,  they do not cover general techniques, research trends, and challenges, which are important aspects in an SLR to drive future activities or the related research area.

To fill these gaps and shed the light on the current state of the applications of EUD for AI, this paper presents a Systematic Literature Review (SLR) that focuses on solutions that support end-users to develop, customize and tailor AI models, and how such activities may shape the future of AI development. In addition, the SLR discusses the current challenges of EUD for AI, the potential benefits, and the future implications of integrating EUD into the overall AI development process.

This paper is structured as follows: Section 2 details the SLR methodology; Section 0 describes the dimensions of the analysis; Section 4 discusses research challenges; Section 5 outlines the threats to the validity of this study, and, finally, Section 6 concludes the article.

## 2     Planning and Conducting the Systematic Literature Review

We conducted a Systematic Literature Review (SLR) via a reproducible and thorough approach to shed the light on the current landscape of EUD for AI systems, i.e., how users, even without skills in AI and/or programming, can customize the AI behavior to their needs. According to Kitchenham, a SLR requires three steps: *planning*, *conducting*, and *reporting* [6]. This section details the first two, while Section 2 details the last one.

### 2.1 Planning the SLR

Planning the SLR includes the following activities [6]: 1) formulation of the research question; 2) definition of the search strings; 3) selection of data sources; 4) definition of inclusion criteria. In the following, we report the details of each activity.

**Formulation of the Research Question.** The main goal of our SLR is to investigate the current state of research on the EUD for AI systems. With this goal in mind, we formulated the following research question: *How users can perform EUD for AI systems?*

Answering this question allows us to provide insights into how the literature tackles the problem of AI customization and democratization. On the other hand, it will drive the identification of future research respectively by focusing on research trends and by presenting the challenges and limitations identified in the available literature.

**Definition of the Search Strings.** We defined a total of 9 search strings by deriving terms from the knowledge of the authors of the subject matter. The strings resulted from a combination of the two keywords "end-user development" and "artificial intelligence". To ensure that most of the literature was covered by our search, we decided to also include the concepts of "no-code" and "low-code" (i.e., names for EUD that are common in modern commercial systems [7-9]), as well as the concept of "customization" (i.e., one of the main goals of EUD, as well as one of our interests). Thus, the resulting strings used to query the search engines were:

- EUD AI
- EUD AI customization
- End-user development AI
- End-user development AI customization
- No-code AI
- No-code AI customization
- Low-code AI
- Low-code AI customization
- AI customization

**Selection of Data Sources.** The chosen search engine is Google Scholar, as it is considered one of the top search engines for scientific researchers and can search the largest databases of scientific publications, ensuring wide coverage of searches.

**Definition of the Inclusion Criteria.** This step concerns the final selection of the relevant publications based on 3 inclusion criteria:

- *Peer reviewed,* i.e., the article is the result of a peer review process. In the case of journal articles, we included publications that appeared in a journal ranked on Scimago as Q2 or Q1, while publications ranked as Q3 were carefully evaluated. In the case of conference articles, we considered the Core Conference Raking, including publications that appeared in a venue with a score of B, A, and A*, while in the case of score C, we carefully evaluated the publication.
- *Written in English*.
- *Focused on EUD and AI.* Relevance to the topic of EUD for AI is assessed by analyzing the title and abstract of each publication, and introduction if needed.

## 2.2   Conducting the Literature Review

After the initial planning phase, the literature review is conducted. Following what Kitchenham suggests, we performed two main activities: the literature review execution, and the data synthesis [6], described in the subsequent subsections.

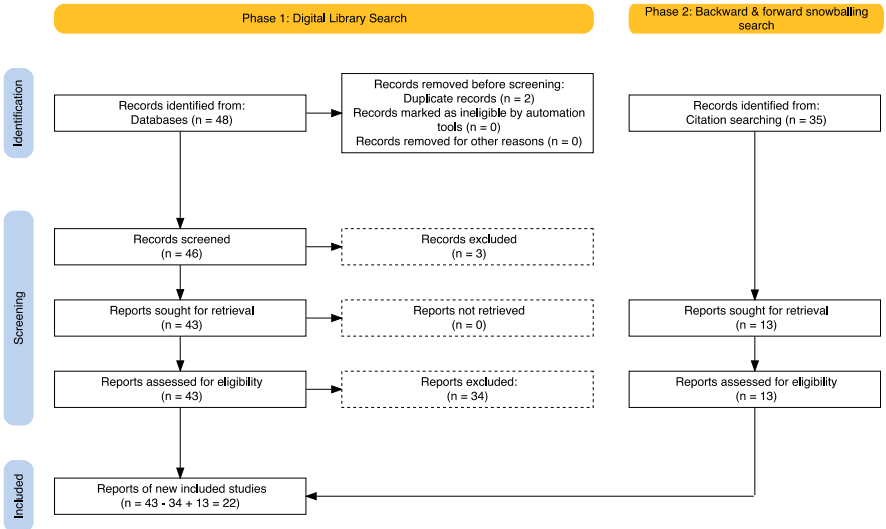

Figure 1. Flow diagram summarizing the selection of the publications along the 2 search phases.

**Literature Review Execution.** This activity was performed from December 2022 – January 2023 following the process depicted in Figure 1, which mainly consists of 2 phases:
- *Phase 1 - Digital library search*: we searched in the Google Scholar digital library using the search strings described in Section 2.1;

- *Phase 2 - Backward and forward snowballing search*: we checked references and citations of the publications resulting from the previous phase, as well as publications that cited publications from Phase 1 [10].

The initial search across the digital library yielded a total of 48 potentially relevant publications. After a check for duplicates, a dataset of 46 publications was obtained. Each publication was then analyzed by reviewing the abstract, the introduction, and the conclusions, considering the inclusion criteria. In the end, this phase resulted in a total of 43 publications. After the application of the inclusion criteria, we excluded 34 publications, thus obtaining a final dataset of 9 publications. Phase 2 allowed us to retrieve further publications, without any time constraint, leading the final set to 22 publications.

**Data Synthesis.** The 22 publications resulting from the search phases are listed in Table 1 and in the References Section. The distribution of the selected publications according to their publication year is shown in Figure 2.

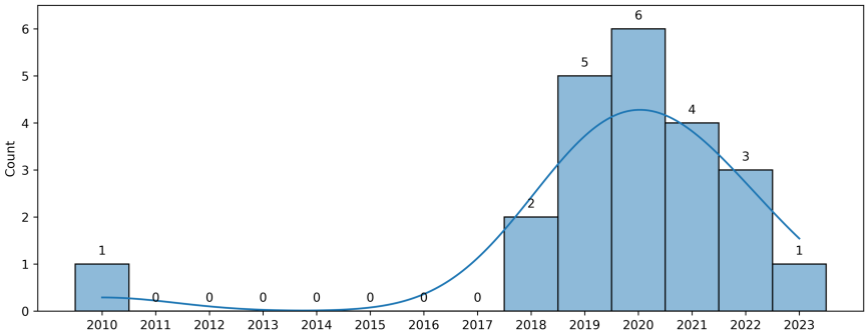

Figure 2. Publications distribution by year.

## 3    Reporting and Analyzing the Results

This section reports the analysis of the literature to answer the research question presented in Section 2.1. Throughout a deep analysis of the selected publications, we defined 8 dimensions that characterize the existing EUD solutions for AI. This provides an overview of the state of the art, but it also provides an initial framework that may aid the design of novel EUD AI-based systems. A summary of the dimensions and the placement of the publications in such dimensions are reported in Table 1. An overview of the distribution of the set of publication in each dimensions category is also shown in Figure 3.

Table 1. Summary of the 8 dimensions and their values for each publication of the SLR.

| Ref. | Composition paradigm | Target users | Technology | Domain | Usage | Customization level | Approach Output |
|---|---|---|---|---|---|---|---|
| [11] | Component-Based | Lay Users | Architecture | AI Model Development | Single | Tailoring | AI Model |
| [12] | Wizard-Based; Rule-Based; Component-Based | Experts | IoT | Education and Teaching | Collaborative | Tailoring | Teaching Suggestions |
| [13] | Template-based | Lay Users; Experts | AI models | Domain-Specific Operations | Collaborative | Customization | AI Model |
| [14] | Rule-Based | Lay Users; Experts | IPA | AI Model Development | Single | Tailoring | AI Model |
| [15] | Component-Based; Workflow and Data Diagrams | Experts | IPA | Interaction Design | Collaborative | Tailoring | AI Model |
| [16] | Template-based | Lay Users | Visual Analytics | Domain-Specific Operations | Collaborative | Customization | Visualization Prototype |
| [17] | Wizard-Based | Lay Users | AI models | Interaction Design | Single | Customization | AI Model |
| [18] | Template-based | Lay Users | AI models | Domain-Specific Operations | Collaborative | Customization | Business Model |
| [19] | Template-based | Lay Users | AI models | Domain-Specific Operations | Collaborative | Customization | Business Model |
| [20] | Component-Based | Lay Users | IoT | Education and Teaching | Single | Customization | AI Model |
| [21] | Component-Based | Lay Users | AI Models | Education and Teaching | Single | Tailoring | AI Model |
| [22] | Component-Based; Workflow and Data Diagrams | Lay Users | AI Models | AI Model Development | Single | Customization | AI Model |
| [23] | Workflow and Data Diagrams | Lay Users | AI Models | Domain-Specific Operations | Single | Tailoring | AI Model |
| [24] | Component-Based | Lay Users | AI Models | AI Model Development | Single | Customization | AI Model |
| [25] | Component-Based[1] | Experts | AI Models | AI Model Development | Single | Creation | AI Model |
| [26] | Component-Based | Lay Users | AI Models | Domain-Specific Operations | Single | Tailoring | AI Model |

---

[1] The publication also presents an approach based on text: for the goal of this study, only the component-based approach is considered interesting.

| Ref. | Composition paradigm | Target users | Technology | Domain | Usage | Customization level | Approach Output |
|---|---|---|---|---|---|---|---|
| [27] | Wizard-Based | Lay Users | AI Models | Domain-Specific Operations | Single | Customization | AI Model |
| [28] | Wizard-Based | Lay Users | Architecture | AI Model Development | Single | Tailoring | AI Model |
| [29] | Component-Based | Experts | IoT | Education and Teaching | Collaborative | Tailoring | Teaching Suggestions |
| [30] | Component-Based | Lay Users; Experts | AI models | Domain-Specific Operations | Collaborative | Customization | AI Model |
| [31] | Wizard-Based | Lay Users; Experts | IPA | AI Model Development | Single | Tailoring | AI Model |
| [32] | Component-Based | Experts | IPA | Interaction Design | Collaborative | Tailoring | AI Model |

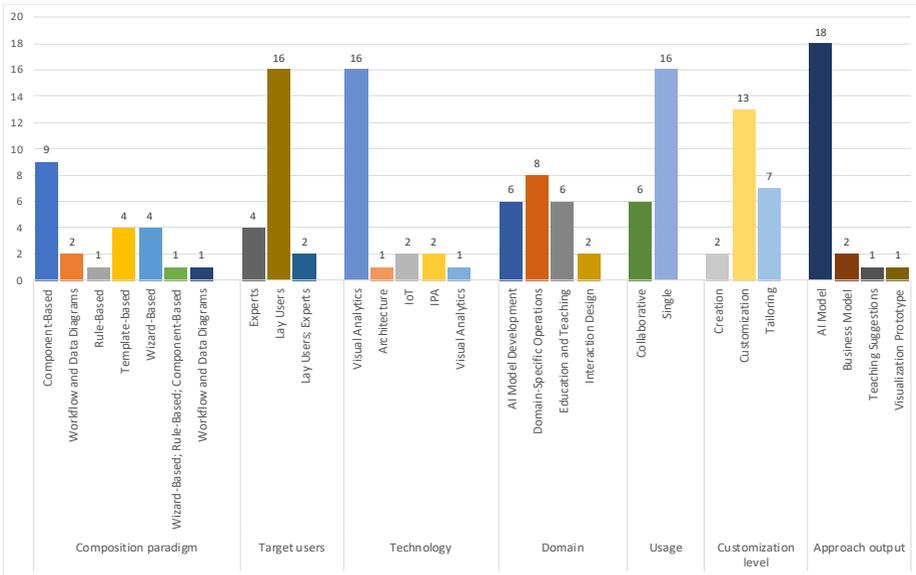

Figure 3. Frequencies of the distributions of the publications in each dimension category.

The resulting dimensions are described in the following paragraphs. For each dimension, we discuss the classification of the publications. Figure 4 provides an overview of the classification framework. It is worth remarking that the dimensions identified for and associated with each publication are not exclusive since a publication can be associated with one or more dimensions.

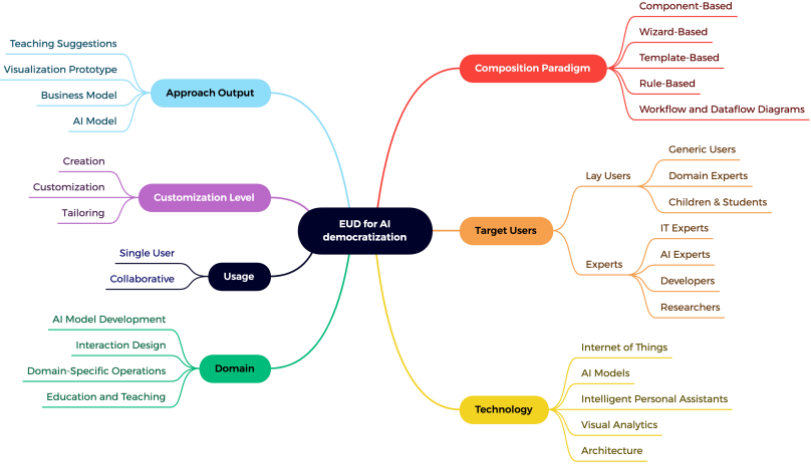

Figure 4. A classification framework for the existing EUD solutions for AI.

**Dimension 1–Composition Paradigm.** EUD researchers proposed several techniques to offer lay users the ability to customize their systems [33]. In the specific context of EUD for AI, we identified 5 techniques that are described in the following paragraphs.

*Component-Based.* It consists of composing 2D or 3D objects that represent domain-specific concepts [33]: a typical example of this technique is the jigsaw metaphor. For example, Piro et al. present an interactive paradigm, which extends an already-existent framework used to build chatbots for conversational AI [15]. The authors' proposal consists in manipulating 2D objects that represent a database's annotation schema and conversation patterns, allowing non-expert programmers to build chatbots from scratch.

*Wizard-Based.* The wizard-based approach is useful in situations where a task can be simplified to a sequence of simple operations that guide the users throughout the overall activity, thus reducing the cognitive load of the task [33]. An example is the proposal by Rodríguez-García et al., who propose a system that guides the users with a wizard-like interface in the process of the training of an AI model (i.e., dataset creation, training, and testing), providing the possibility of using the trained model in other EUD tools (e.g., Scratch) [27].

*Template-Based.* It consists of presenting to the end-users pre-made and customizable functionalities, allowing them to edit parameters and/or text to fit their own needs [33]. For example, Iyer et al. present *Trinity*, a system for Data Mining that enables the end-users to visually perform three tasks: experiment management, label management, and feature management. For each task, this tool proposes a template with a set of default configurations that can simply be modified by the end users [13].

*Rule-Based.* It allows end-users to tailor AI components by defining trigger-actions rules for specific purposes [33]. For instance, Rough et al. identify and categorize existing EUD for AI tools applied to the Intelligent Personal Assistant (IPA) domain: they illustrate the opportunity given to end-users in creating their own rules to customize skills or construct routines by composing a rule, after the identification of trigger events and subsequent actions [14].

*Workflow and Dataflow Diagrams.* This technique concerns the way data are manipulated in the proposed solutions. It consists of using graphical elements, which can be interconnected and intertwined, depending on the needs of the users (i.e., a graph) [33]. The approach can be useful for programmers and developers who do not have a professional or technical background by enabling them to have a visual way to manipulate workflows and data. For example, Godec et al. propose a solution that allows the definition of a pipeline to automatically analyze medical images [23].

**Dimension 2–Target Users.** One of the key requirements in designing an EUD tool for AI is to carefully identify who the end users will be, considering both their limitations and strengths. The main goal is to facilitate their experience and use of the final tool. The analyzed publications led to the identification of 4 main target users, which are described in the following paragraphs.

*Lay Users.* They can be defined as people who lack any IT or AI-related skills. In this class, we identified generic users and domain experts. For this review, we consider children and students also as lay users (regardless of their education in IT) as they are not IT or AI experts. Most of the retrieved research focuses on this class of users. For example, Zimmermann-Niefield et al. propose an application that allows its users to train and use ML models that help them during athletic exercises, without requiring any IT or AI knowledge [31]. Another example focused on domain experts is the proposal by Sanctorum et al., who strive to find solutions to help subject matter experts autonomously manage a knowledge base in the toxicology domain [11].

*Experts.* This category includes users that are experts in IT and/or AI. Here we identified different types of experts, namely *IT experts*, *AI experts*, *developers,* and *researchers*. *IT experts* are users that possess a technical background in IT but do not have any AI or developer skills. Kahn et al. provide a case study, showing that a block-based programming environment may be better for this kind of user [24]. On the other hand, for AI experts we mean users that are experts both in the field of IT and AI, such as AI Specialists and Machine Learning Experts. For instance, Shaikh provides an example of low-code AI customization, geared toward experts in IT that know AI models to choose the best Microsoft Azure services to create an AI-based system [28]. By *developers* we mean users who are capable of programming: for example, Rough and Cowan discuss how the majority of EUD solutions for personal assistants are targeted toward developers, who know how to use Software Development Kits [14]. The last category is *researchers,* i.e., users that adopt this kind of solution for research purposes. For example, Tamilselvam et al. present and test a solution that allows the quick prototyping of neural networks using either tables or a drag-and-drop interface, allowing researchers to quickly create neural networks for further usage [29].

**Dimension 4–Technology.** The technology defines the constraints, the possibilities, and the context of the use of the proposed solutions. In this context, we identified 5 different types of technologies: the Internet of Things (IoT), AI Models, Intelligent Personal Assistants (IPAs), Visual Analytics, and Architecture.

*Internet of Things.* The Internet of Things (IoT) is defined as "An open and comprehensive network of intelligent objects that can auto-organize, share information, data, and resources, reacting and acting in face of situations and changes in the environment" [34]. We classify in this category the research that was focused mainly on IoT itself, discussing AI as a secondary subject. For example, Agassi et al. discuss an IoT system that recognizes gestures, and they provide a solution that allows users to customize the recognition model itself [20].

*AI Models.* We classify as part of this category all research that is mainly focused on the customization, tailoring, or creation of the AI models themselves, regardless of the specific context of use. For example, Xie et al. present an IDE plugin that allows its users to create, through the paradigms of visual programming, and visualize the neural network they are building [30].

*Intelligent Personal Assistant.* An *IPA* is defined as a user interface, which main interaction method is through speech, that aims at performing tasks requested in natural language [14]. We classify in this category all publications that aim at enabling users to customize chatbots or personal assistants. For example, Rough and Cowan highlight existing EUD opportunities to allow users to customize their personal assistants to make them truly "personal" [14].

*Visual Analytics.* Visual analytics technology supports the analysis of datasets through sophisticated tools and processes using visual and graphical representations. An example is the solution proposed by Mishra et al., which discuss the development of a prototype tool that enables leaderboard revamping through customization, based on the focus area that the end-user is interested in [16].

*Architecture.* The last category does not refer to a concrete technology but to an abstract architecture to design EUD systems for AI. For instance, Sanctorum et al. propose an architecture representing the key phases and tasks of knowledge graphs' lifecycle, to guide end-users in the definition of custom knowledge bases [11].

**Dimension 5–Domain.** With this dimension, we aim at classifying the focus of each publication. The domain of the publication sets its goal, and it may heavily influence the approach used in the publications. A total of 4 dimensions have been identified.

*AI Model Development.* It represents the class of publications that mainly aim at providing the end-users with a customized AI model that they can use for either a generic task (chosen by them during the customization) or for a predefined task. For example, Carney et al. propose a solution that allows end users to train, through the manipulation of graphical elements, a pre-defined neural network architecture for classification, leaving the users the freedom to choose the classes themselves [22].

*Interaction Design.* We classify in this domain all the publications that aim at defining the way the interactive system supports users' activities and interacts with them [35]. For example, Bunt et al. try to discover new ways of interacting with AI-based systems, specifically with an AI-enabled recommender system-like tool, which can provide users with suggestions on how to customize graphical interfaces according to their personal preferences [17].

*Domain-Specific Operations.* This general domain comprises all publications that aim at allowing users to customize or create AI models that are designed to perform a specific task in a specific domain. For example, Jauhar et al. explore how EUD and AI can be used in inventory and supply chains to enable retailers and operators to manage, most coherently and consistently as possible, managing costs, processes, anomalies, and predictions by using machine learning algorithms [36].

*Education and Teaching.* The last domain identified is the one of education and teaching; it concerns how EUD, with its different subtopics, can be taught in various contexts and how awareness can be spread among developers, designers, and engineers. An example is by Paternò, who explores how AI and IT experts need to empower and encourage end users in using and creating daily automations [12].

**Dimension 6–Usage.** The *Usage* dimension regards the number of people that the system allows to collaborate during its employment.

*Single User.* It refers to the use of a system, or a solution, by a single individual. For example, Sanctorum et al. present an approach to let the subject-matter expert alone manually manage a toxicology knowledge base [11].

*Collaborative.* It refers to the use of a system, or a solution, by a group of individuals with multiple perspectives and experiences. For instance, Iyer et al. propose a system that holds a shared vocabulary and workspace, enabling better collaboration between more end users while recalibrating their skills as if they were equal partners [13].

**Dimension 7–Customization Level.** This dimension refers to the level of allowed modifications to the AI-based software's components. We identify three possibilities: creation, customization, and tailoring.

*Creation.* Systems that allow the "creation" of AI models are systems that do not pose any limitation to the personalized system capabilities, nor do they provide a pre-made system that can be adapted to the users' Needs. Examples are by Xie et al. and Tamilselvam et al., who both define solutions that allow end users to graphically build neural networks [29, 30].

*Customization.* With customization, we identify all the approaches that allow the users to heavily alter the parameters of an existing AI model. However, the coarse task (e.g., classification) and the model architecture are predefined. For example, the solution proposed by Carney, et al. allows training a pre-defined neural network for classification, providing the ability to customize the dataset and the available labels [22].

*Tailoring.* We categorize as "tailoring" systems, all solutions that allow end users to *fine-tune* a model for their specific needs, even though the specific task is predetermined. An example is the gesture recognition system proposed by Agassi et al., in which users can change specific gestures, but they are not able to edit other aspects of the AI model [20].

**Dimension 8–Approach Output.** This dimension was identified to define the types of results and outcomes of the proposed solutions. A total of 4 categories were identified.

*Teaching Suggestions.* Discussion concerning how EUD should be thought about AI is still open. An example of the discussion is in the work by Paternò, who provides suggestions as to how experts should encourage end users in using and creating daily automations [12].

*Visualization Prototype.* EUD solutions for AI customization may also aid the users by presenting results and metrics in different modalities. For example, Mishra et al. propose visualization techniques that aid the end users in selecting models that are trained and compared fairly [16].

*Business Model.* The adoption of EUD may aid in undertaking business decisions and may provide new business models to be exploited. This is suggested, for example, by Redchuk et al., who show that adopting no/low-code AI technologies may shorten the implementation cycles of new systems [18].

*AI Model.* The main output type concerns the AI model itself. We classify in this cateogry all the approaches that aim at providing the end users with their customized AI model. Most of the publications provide this type of output: for example, Piro et al. allow users to define chatbots that can be deployed immediately [15]. Similarly, Tamilselvam et al. propose a method that allows users to graphically build neural networks for later use [29].

## 4    Future Challenges

Throughout this SLR, some interesting and relevant challenges emerged that are worth reporting. This section thus discusses relevant open questions that may guide future research in EUD for AI.

**Adopt a Human-Centered Artificial Intelligence approach.** In recent years, a new perspective has emerged that aims to reconsider the centrality of humans while reaping the benefits of AI systems to augment rather than replace professional skills: Human-Centered AI (HCAI) is a novel framework that posits that high levels of human control are not incompatible with high levels of computer automation [37]. EUD for AI will play a central role in this direction. As emphasized by Schmidt, AI-based systems should allow humans to have an appropriate level of control by providing EUD approaches for the initial configuration of the system, as well as for reconfiguration of the system at the time of use, to satisfy user needs that cannot be anticipated by the automated system [38]. This is an approach that has not been much explored in the literature, so we believe that EUD can contribute in this direction.

**Support collaborative activities.** Another challenge lies in the collaborative aspect of the EUD for AI. Very little work is available on this aspect: only 6 publications out of the retrieved 22 acknowledge the collaborative aspect of AI development. However, the creation, testing, and deployment of AI models (and AI systems in general) is a collaborative activity that involves multiple actors with different expertise [39]. This highlights the requirement of collaboration in no/low-code AI tools for them to be used in real-case scenarios, especially by experts.

**Provide EUD solutions also for AI experts**. The SLR found that most of the research is focused on lay users, while a few studies target experts. Although EUD for AI could lead to a democratization of AI technologies, it may be interesting to better explore how AI experts can benefit from EUD solutions for AI. In fact, although experts might create or adapt AI systems using technical tools (e.g., Python, R, Weka, etc.), providing an EUD environment might allow them to optimize their resources [18]. An example is Knime, an end-to-end data science platform that provides a graphical user interface based on the graph metaphor to support experts from building analytical models to deploying them without coding [40].

**Enable the creation of AI solutions through EUD**. A strong limitation of the current research on this topic is the scarce focus on AI system *creation* rather than *customization* or *tailoring*: only 2 out of 22 retrieved publications reach this goal. Furthermore, these 2 publications focus on expert users: research on AI model creation by lay users is, as far as we could find out, completely missing. The lack of the possibility for end users to create rather than adapt the model also affects solutions aimed at experts [27].

**Define the right composition paradigm for each type of user.** At the time of the review, no studies were available on the relationship between EUD programming paradigms and the level of expertise and skills of the users. This is a very important aspect that has been addressed in a similar context. For example, trigger-action programming of IoT devices was found to be easier for non-technical users when using wizard procedures as well as component- or template-based paradigms, while graph-based metaphor was found to be more effective for technical users [41, 42]. Certainly, lessons learned in similar domains can be a good starting point for EUD for AI, but further research is needed to provide the users with the right abstraction mechanisms.

## 5    Threats to validity

Several threats to validity can affect the results of a systematic literature review. In the following, we report how we mitigated the most critical ones.

**Selection bias**. This occurs when the studies included in the review are not representative of the entire population of studies on the topic. This has been mitigated by i) manually reviewing the publication to ensure their compliance with the SLR goal, and ii) performing two phases, i.e., search on digital libraries and snowballing.

**Publication bias**. This occurs when studies that show statistically significant results are more likely to be published than studies that do not. This aspect has been mitigated by manually reading those publications that do not report any result but only a technical solution with preliminary results. Besides the generic inclusion criteria, their relevance for our SLR considered, for example, the number of citations and the novelty of the solution.

**Time lag bias**. This occurs when the review does not include all relevant studies because were published after the review was conducted. In this case, we can safely assume that this threat is not so evident in our study since the SLR has been performed 2 months before its submission.

**Publication quality**. This occurs when studies of poor quality are included in the review. To mitigate this aspect, we defined inclusion criteria on the quality of the venue of the publication, leaving to a manual evaluation of the authors of this SRL the inclusion of publications that appeared in venues of lower quality.

## 6     Conclusions

This SLR provides an overview of the current state of research in the field of EUD for no/low-code AI creation and customization. The first contribution is the identification of the main topics that are discussed in the community, highlighting potential benefits that lay users or businesses may get from the adoption of EUD technologies. A second contribution, the SLR sheds the light on existing limitations and key challenges that affect the current research landscapes, identifying some insights into possible future research directions. In future work, this SLR might be extended in different directions. First, further digital libraries might be considered, for example, Scholar, Elsevier, and IEEE. Then, other keywords might be defined to increase coverage. Finally, an additional manual search can be performed on Conferences Proceedings and Journals relevant to the topics of the SLR.